\documentclass[12pt]{article}%
\usepackage{amsmath}
\usepackage{amsfonts}
\usepackage{amssymb}
\usepackage{graphicx}%
\providecommand{\U}[1]{\protect\rule{.1in}{.1in}}
\providecommand{\U}[1]{\protect\rule{.1in}{.1in}}

\begin{document}
\begin{titlepage}
\vspace{.3cm} \vspace{1cm}
\begin{center}
\baselineskip=16pt \centerline{\Large\bf Inhomogeneous Dark Energy }
\vspace{2truecm}
\centerline{\large\bf Ali H.
Chamseddine$^{1,2}$\ , \ Viatcheslav Mukhanov$^{3,4}$\ \ } \vspace{.5truecm}
\emph{\centerline{$^{1}$Physics Department, American University of Beirut, Lebanon}}
\emph{\centerline{$^{2}$I.H.E.S. F-91440 Bures-sur-Yvette, France}}
\emph{\centerline{$^{3}$Theoretical Physics, Ludwig Maxmillians University,Theresienstr. 37, 80333 Munich, Germany }}
\end{center}
\vspace{2cm}
\begin{center}
{\bf Abstract}
\end{center}
We modify Einstein General Relativity
by adding non-dynamical scalar fields to account simultaneously for both dark matter and dark energy.
The dark energy in this case can be distributed in-homogeneously even within horizon scales. Its inhomogeneities can contribute
to the late time integrated Sachs-Wolfe effect, possibly removing some of the low multipole anomalies in the temperature fluctuations
of the CMB spectrum.
The presence of the inhomogeneous dark matter also influences structure formation in the universe.
\end{titlepage}

In the paper $\cite{mimetic}$ we have suggested a minimal extension of General
Relativity where the scalar mode of the metric becomes dynamical even in the
absence of the usual matter. This extra gravitational degree of freedom mimics
cold dark matter. The simplest way to describe it is to introduce the scalar
field $\phi$ satisfying the consrtaint
\begin{equation}
g^{\mu\nu}\partial_{\mu}\phi\partial_{\nu}\phi=1,\label{unitlength}%
\end{equation}
and to implement this consrtaint in the Einstein action via a Lagrange
multiplier \cite{Golovnev}, \cite{Barvinsky}, \cite{mimeticcos}%
\begin{equation}
S=%
%TCIMACRO{\dint }%
%BeginExpansion
{\displaystyle\int}
%EndExpansion
d^{4}x\sqrt{-g}\left[  -\frac{1}{2}R\left(  g_{\mu\nu}\right)  +\lambda\left(
g^{\mu\nu}\partial_{\mu}\phi\partial_{\nu}\phi-1\right)  \right]  ,\label{1}%
\end{equation}
where we set $8\pi G=1.$ In this paper we will show how to extend this model,
in a minimal way, to imitate space-dependent dark energy. The scalar field in
the slow roll regime can also imitate the space-time dependent dark energy but
this field is necessarily dynamical and finally decays. Moreover, small
inhomogenieties in the scalar field cannot be amplified on scales smaller than
the Jeans wavelength, which is normally about the curvature scale. Therefore,
if the presently observed dark energy is due to the dynamical scalar field
then it is homogeneous within present horizon scales. In this paper we
consider non-dynamical scalar field which can imitate arbitrarily
inhomogeneous dark energy in any scale and can contribute to the gravitational
instability and late time integrated Sachs-Wolfe effect at the stage of the
galaxy formation. Let us consider the theory with action given by%
\begin{equation}
S=%
%TCIMACRO{\dint }%
%BeginExpansion
{\displaystyle\int}
%EndExpansion
d^{4}x\sqrt{-g}\left[  -\frac{1}{2}R\left(  g_{\mu\nu}\right)  +\lambda\left(
g^{\mu\nu}\partial_{\mu}\phi\partial_{\nu}\phi-1\right)  +\lambda_{a}g^{\mu
\nu}\partial_{\mu}\phi^{a}\partial_{\nu}\phi-V\left(  \phi^{a}\right)
\right]  \label{3}%
\end{equation}
where $\phi^{a}$ are the extra scalar fields, the number of which is
irrelevant for our considerations. We take the potential $V$ to be a function
of $\phi^{a}$ only, although we could add to it a part depending on $\phi$
only as was done in our previous work \cite{mimeticcos}. The simplest case is
to take one field $\phi^{a}$ only with $a=1$, however, the case of three
scalars $a=1,2,3,$ correspond with identifying the scalars with the
synchronous coordinates. Variations with respect to the Lagrange multipliers
give the following constraints%
\begin{equation}
\frac{\partial\phi}{\partial x^{\mu}}\frac{\partial\phi}{\partial x^{\nu}%
}g^{\mu\nu}=1,\text{ \quad\ }\frac{\partial\phi^{a}}{\partial x^{\mu}}%
\frac{\partial\phi}{\partial x^{\nu}}g^{\mu\nu}=0,\text{\ \ }\label{4}%
\end{equation}
which can be easily solved in the synchronous coordinate system where
$g_{00}=1,$ $g_{0i}=0,$ $i=1,2,3$ so that \cite{Landau}%
\begin{equation}
ds^{2}=dt^{2}-\gamma_{ik}dx^{i}dx^{k}.\label{4a}%
\end{equation}
The general solutions of equations $\left(  \ref{4}\right)  $ in this
coordinate system are%
\begin{equation}
\phi=t,\quad\text{ }\phi^{a}=\phi^{a}\left(  x^{i}\right)  .\label{5}%
\end{equation}
Thus the fields $\phi^{a}$ are time independent and the potential in $\left(
\ref{3}\right)  $ can be an arbitrary time independent function of the spacial
coordinates in the synchronous coordinate system. This potential leads to
cosmological-like constant in the Einstein equations. However, it cannot just
be an inhomogeneous cosmological constant and therefore the effective energy
momentum tensor for the inhomogeneous field $\phi^{a}$ must be different from
those ones which correspond to the standard cosmological constant.

Variation of action $\left(  \ref{3}\right)  $ with respect to the metric
$g_{\mu\nu}$ leads to the following modified Einstein equations%
\begin{equation}
G_{\nu}^{\mu}=\bar{T}_{\nu}^{\mu},\label{6}%
\end{equation}
where
\begin{equation}
\bar{T}_{\nu}^{\mu}=2\lambda\partial^{\mu}\phi\partial_{\nu}\phi+\lambda
_{a}\left(  \partial^{\mu}\phi^{a}\partial_{\nu}\phi+\partial_{\nu}\phi
^{a}\partial^{\mu}\phi\right)  +\delta_{\nu}^{\mu}V\left(  \phi^{a}\right)
,\label{7}%
\end{equation}
is the contribution of mimetic fields $\phi$ and $\phi^{a}.$ The equations for
the Lagrange multipliers $\lambda$ and $\lambda_{a}$ follow from the action
$\left(  \ref{3}\right)  $ under variation with respect to $\phi$ and
$\phi^{a}$%
\begin{equation}
\partial_{\mu}\left(  \sqrt{-g}\left(  2\lambda g^{\mu\nu}\partial_{\nu}%
\phi+\lambda_{a}g^{\mu\nu}\partial_{\nu}\phi^{a}\right)  \right)  =0,\label{8}%
\end{equation}%
\begin{equation}
\partial_{\mu}\left(  \sqrt{-g}\lambda_{a}g^{\mu\nu}\partial_{\nu}\phi\right)
=-\sqrt{-g}V_{,a},\label{9}%
\end{equation}
where $V_{,a}=\frac{\partial V}{\partial\phi^{a}}.$ In the synchronous
coordinate system these equations simplify to
\begin{equation}
\partial_{0}\left(  2\sqrt{\gamma}\lambda\right)  =\partial_{i}\left(
\sqrt{\gamma}\lambda_{a}\gamma^{ik}\partial_{k}\phi^{a}\right)  ,\label{10}%
\end{equation}%
\begin{equation}
\partial_{0}\left(  \sqrt{\gamma}\lambda_{a}\right)  =-\sqrt{\gamma}%
V_{,a},\label{11}%
\end{equation}
and can be easily integrated to give%
\begin{equation}
\lambda_{a}=-\frac{1}{\sqrt{\gamma}}\int^{t}V_{,a}\sqrt{\gamma}dt^{\prime
},\label{12}%
\end{equation}%
\begin{equation}
\lambda=-\frac{1}{2\sqrt{\gamma}}\int^{t}\partial_{i}\left(  \gamma
^{ik}\partial_{k}V\int^{t^{\prime}}\sqrt{\gamma}dt^{\prime\prime}\right)
dt^{\prime}\label{13}%
\end{equation}
Taking this into account we find the following components of the energy
momentum tensor for mimetic matter%
\begin{equation}
\bar{T}_{0}^{0}=-\frac{1}{\sqrt{\gamma}}\int^{t}\partial_{i}\left(
\gamma^{ik}\partial_{k}V\int^{t^{\prime}}\sqrt{\gamma}dt^{\prime\prime
}\right)  dt^{\prime}+V,\label{14}%
\end{equation}%
\begin{equation}
\bar{T}_{i}^{0}=-\frac{1}{\sqrt{\gamma}}\int^{t}\partial_{i}V\sqrt{\gamma
}dt^{\prime},\text{ \quad\ }\bar{T}_{k}^{i}=V\delta_{k}^{i}.\label{15}%
\end{equation}
One can easily verify that $\bar{T}_{\beta}^{\alpha}$ given above
automatically satisfy the conservation laws $\nabla^{\beta}\bar{T}_{\beta
}^{\alpha}=0,$ as it must be. To clarify the meaning of the constants of
integration in $\left(  \ref{14}\right)  $ and $\left(  \ref{15}\right)  $ it
is convenient to consider the case of vanishing potential,$\ V=0,$ and find
directly the solution of equations $\left(  \ref{8}\right)  $ and $\left(
\ref{9}\right)  $:%
\begin{equation}
\lambda_{a}=\frac{g_{a}\left(  x^{k}\right)  }{\sqrt{\gamma}},\quad\text{
}\lambda=\frac{C\left(  x^{k}\right)  }{2\sqrt{\gamma}}+\frac{1}{2\sqrt
{\gamma}}\int^{t}\partial_{i}\left(  \gamma^{ik}g_{a}\partial_{k}\phi
^{a}\right)  dt^{\prime},\label{16}%
\end{equation}
where it is assumed that the integral is equal to zero in case of vanishing
integrand$.$ The corresponding energy momentum tensor is%
\begin{equation}
\bar{T}_{0}^{0}=\frac{C\left(  x^{k}\right)  }{\sqrt{\gamma}}+\frac{1}%
{\sqrt{\gamma}}\int^{t}\partial_{i}\left(  \gamma^{ik}g_{a}\partial_{k}%
\phi^{a}\right)  dt^{\prime},\label{17}%
\end{equation}%
\begin{equation}
\bar{T}_{i}^{0}=\frac{g_{a}\partial_{i}\phi^{a}}{\sqrt{\gamma}},\text{ \quad
}\bar{T}_{k}^{i}=0.\label{18}%
\end{equation}
It is clear that the first term in $\left(  \ref{17}\right)  $ describes
mimetic dark matter. To figure out what kind of extra degrees of freedom
induces the term proportional to $g_{a}\partial_{i}\phi^{a}$ let us consider
perturbations in a flat Friedmann universe assuming that $g_{a}\partial
_{i}\phi^{a}$ is small. In the synchronous coordinate system the metric of a
slightly perturbed universe takes the form \cite{Mbook}:%
\begin{equation}
ds^{2}=dt^{2}-a^{2}\left(  t\right)  \left[  \left(  \left(  1-2\psi\right)
\delta_{ik}-2E_{,ik}-F_{i,k}-F_{k,i}\right)  dx^{i}dx^{k}\right]  ,\label{19}%
\end{equation}
where $\psi$ and $E$ describe the scalar metric perturbations and $\ F_{i}$
the vector perturbations which satisfies $F_{i,i}=0$. We have also skipped the
gravitational waves. Assuming that $C\left(  x^{k}\right)  =C_{0}+\delta
C\left(  x^{k}\right)  ,$ decomposing $g_{a}\partial_{i}\phi^{a}$ into
longitudinal and transverse parts as%
\[
g_{a}\left(  x^{k}\right)  \partial_{i}\phi^{a}\left(  x^{i}\right)  =\left(
\partial_{i}\sigma+\sigma_{i}^{T}\right)  ,\quad\text{ }\sigma_{i,i}^{T}=0
\]
and taking into account that to linear order in perturbations%
\begin{equation}
\frac{1}{\sqrt{\gamma}}=\frac{1}{a^{3}}\left(  1+3\psi+\Delta E\right)
,\label{20}%
\end{equation}
we find
\begin{equation}
\bar{T}_{0}^{0}=\frac{C_{0}}{a^{3}}+\frac{\delta C}{a^{3}}+\frac{C_{0}}{a^{3}%
}\left(  3\psi+\Delta E\right)  +\frac{\Delta\sigma}{a^{3}}\int\frac{dt}%
{a^{2}},\label{21}%
\end{equation}%
\begin{equation}
\bar{T}_{i}^{0}=\frac{\left(  \partial_{i}\sigma+\sigma_{i}^{T}\right)
}{a^{3}}\label{22}%
\end{equation}
We will now show that the longitudinal part of perturbations due to $\sigma$
can be removed by a coordinate transformation which preserves the synchronous
coordinate system. The general infinitesimal coordinate transformations can be
written as
\begin{equation}
t\rightarrow\tilde{t}=t+\xi^{0}\left(  x^{\alpha}\right)  ,\text{ }%
x^{i}\rightarrow\tilde{x}^{i}=x^{i}+\zeta_{,i}\left(  x^{\alpha}\right)
+\xi_{T}^{i}\left(  x^{\alpha}\right)  ,\label{23}%
\end{equation}
where $\xi_{T,i}^{i}=0.$ Requiring that $g_{00}\rightarrow\tilde{g}_{00}=1$
and $g_{0i}\rightarrow\tilde{g}_{0i}=0$ we find that only the coordinate
transformations with%
\begin{equation}
\xi^{0}=\xi^{0}\left(  x^{i}\right)  ,\text{ \ }\zeta=\xi^{0}\left(
x^{i}\right)  \int\frac{dt}{a^{2}},\text{ \ }\xi_{T}^{i}=\xi_{T}^{i}\left(
x^{i}\right)  ,\label{24}%
\end{equation}
stay within the synchronous coordinate system. Under transformations $\left(
\ref{24}\right)  $ the metric perturbations change according to
\begin{equation}
\psi\rightarrow\tilde{\psi}=\psi+\frac{\dot{a}}{a}\xi^{0},\text{ }%
E\rightarrow\tilde{E}=E+\xi^{0}\int\frac{dt}{a^{2}},\text{ \ }F_{i}%
\rightarrow\tilde{F}_{i}=F_{i}+\xi_{T}^{i},\label{25}%
\end{equation}
where dot denotes the derivative with respect to time $t.$ Noting that
$\bar{T}_{i}^{0}$ components are transformed as
\begin{equation}
\bar{T}_{i}^{0}\rightarrow\bar{T}_{i}^{0}-\xi_{,i}^{0}\bar{T}_{0}^{0}%
=\frac{\left(  \partial_{i}\sigma+\sigma_{i}^{T}\right)  }{a^{3}}-\xi_{,i}%
^{0}\frac{C_{0}}{a^{3}}\label{26}%
\end{equation}
we take $\xi^{0}=\sigma/C_{0}.$ As a result the $\bar{T}_{i}^{0}$ components
in the new coordinate system become%
\begin{equation}
\widetilde{\bar{T}}_{i}^{0}=\frac{\sigma_{i}^{T}}{a^{3}}.\label{27}%
\end{equation}
Correspondingly
\begin{equation}
\widetilde{\bar{T}}_{0}^{0}=\frac{C_{0}}{a^{3}\left(  \tilde{t}\right)
}+\frac{\delta C}{a^{3}}+\frac{C_{0}}{a^{3}}\left(  3\tilde{\psi}+\Delta
\tilde{E}\right)  =\frac{C\left(  x^{k}\right)  }{\sqrt{\tilde{\gamma}}%
}.\label{28}%
\end{equation}
To derive this formula we took into account that at a given point of the
manifold $\widetilde{\bar{T}}_{0}^{0}=\bar{T}_{0}^{0}$ and then replaced in
$\left(  \ref{21}\right)  $ $\sigma$ by $C_{0}\xi^{0}.$ After that the first
term on the right hand side was expanded as
\[
a\left(  t\right)  =a\left(  \tilde{t}-\xi^{0}\right)  =a\left(  \tilde
{t}\right)  -\dot{a}\left(  t\right)  \xi^{0},
\]
and finally we have used $\left(  \ref{25}\right)  .$ Thus, in the absence of
the potential $V\left(  \phi^{a}\right)  $ the extra fields $\phi^{a}$ do not
produce additional longitudinal perturbations in the mimetic matter but allow
us to describe the vector (rotational) perturbations of the mimetic dust. As a
result the mimetic matter becomes completely indistinguishable from dust in
the linearized approximation. \ 

Now let us consider the behavior of the extra modes due to $\phi^{a}$ in the
presence of the potential $V\left(  \phi^{a}\right)  .$ Neglecting the metric
perturbations the components of the energy momentum tensor $\left(
\ref{14}\right)  ,\left(  \ref{15}\right)  $ in the Friedmann universe become%
\begin{equation}
\bar{T}_{0}^{0}=-\frac{\Delta V}{a^{3}\left(  t\right)  }\int^{t}\left(
\int^{t^{\prime}}a^{3}\left(  t^{\prime\prime}\right)  dt^{\prime\prime
}\right)  \frac{dt^{\prime}}{a^{2}\left(  t^{\prime}\right)  }+V,\label{29}%
\end{equation}%
\begin{equation}
\bar{T}_{i}^{0}=-\frac{\partial_{i}V}{a^{3}\left(  t\right)  }\int^{t}%
a^{3}\left(  t^{\prime}\right)  dt^{\prime},\text{ \quad\ }\bar{T}_{k}%
^{i}=V\delta_{k}^{i}.\label{30}%
\end{equation}
If the fields $\phi^{a}$ and $V$ are inhomogeneous, then the energy momentum
tensor looks as if describing the inhomogeneous \textquotedblleft cosmological
constant\textquotedblright. In fact, assuming that the random field $\phi^{a}$
is isotropic after averaging over large scales we find
\begin{equation}
\bar{T}_{0}^{0}=\left\langle V\right\rangle ,\quad\text{ }\bar{T}_{k}%
^{i}=\left\langle V\right\rangle \delta_{k}^{i},\text{ \quad}\bar{T}_{i}%
^{0}=0,\label{31}%
\end{equation}
where $\left\langle V\right\rangle $ denotes the spacial average of the
potential over a scale exceeding the scales of inhomogenieties. Thus
irrespective of how clumpy is the distribution of $V\left(  x^{i}\right)  ,$
it contributes to the large scale evolution of the universe exactly as a
cosmological constant.

To clarify how inhomogenieties of $\phi^{a}$ behave in a flat Friedmann
universe we will consider small perturbations $\delta\phi^{a}$ of the
homogeneous background $\phi_{0}^{a}=\mathrm{const}$ in the
conformal-Newtonian coordinate system, where the metric takes the form%
\begin{equation}
ds^{2}=\left(  1+2\Phi\right)  dt^{2}-a^{2}\left(  t\right)  \left(
1-2\Phi\right)  \delta_{ik}dx^{i}dx^{k}.\label{32}%
\end{equation}
Here we have used $\delta T_{k}^{i}$ $=0$ for $i\neq k$ to equate two
gravitational potentials \cite{Mbook}. The solution of equations $\left(
\ref{4}\right)  ,\left(  \ref{8}\right)  $ and $\left(  \ref{9}\right)  $ for
the homogeneous background is%
\begin{equation}
\phi=t,\quad\text{ }\phi^{a}=\phi_{0}^{a},\text{ \quad}\lambda_{0}=\frac
{C_{0}}{2a^{3}\left(  t\right)  },\quad\text{ }\lambda_{a0}=-\frac{V_{,a}%
}{a^{3}\left(  t\right)  }\int^{t}a^{3}\left(  t^{\prime}\right)  dt^{\prime
},\label{33}%
\end{equation}
where $\phi_{0}^{a}$ and $C_{0}$ are constants which do not depend on space
and time. Considering small perturbations about a homogeneous background, we
find to linear order%
\begin{equation}
\delta\dot{\phi}=\Phi,\text{ \quad}\delta\phi^{a}=\delta\phi^{a}\left(
x^{i}\right)  ,\text{ }\label{34}%
\end{equation}%
\begin{equation}
\delta\lambda=3\lambda_{0}\Phi+\frac{1}{a^{3}\left(  t\right)  }\int%
^{t}a\lambda_{0}\Delta\delta\phi dt^{\prime}+\frac{\Delta\delta\phi^{a}%
}{2a^{3}\left(  t\right)  }\int^{t}\lambda_{a0}adt^{\prime},\label{35}%
\end{equation}
where we have denoted by dot derivative with respect to time $t.$ The
corresponding perturbations of the energy momentum tensor for mimetic matter
in this coordinate system are%
\begin{equation}
\delta\bar{T}_{0}^{0}=2\delta\lambda+V_{,a}\delta\phi^{a},\quad\delta\bar
{T}_{i}^{0}=\frac{1}{a}\partial_{i}\left(  2\lambda_{0}\delta\phi+\lambda
_{a0}\delta\phi^{a}\right)  ,\text{ \ }\delta\bar{T}_{i}^{k}=V_{,a}\delta
\phi^{a}\delta_{i}^{k}\label{36}%
\end{equation}
The $0-i$ Einstein equation for perturbations reduces to
\begin{equation}
\dot{\Phi}+H\Phi=\lambda_{0}\delta\phi+\frac{1}{2}\lambda_{a0}\delta\phi
^{a}\label{37}%
\end{equation}
where $H=\dot{a}/a.$ Taking into account that $\dot{H}=-\lambda_{0}$ and
$\Phi=\delta\dot{\phi}$, this equation becomes%
\begin{equation}
\delta\ddot{\phi}+H\delta\dot{\phi}+\dot{H}\delta\phi=\frac{1}{2}\lambda
_{a0}\delta\phi^{a}\label{38}%
\end{equation}
and can be easily integrated to give
\begin{equation}
\delta\dot{\phi}+H\delta\phi=\frac{1}{2}\delta\phi^{a}\int^{t}\lambda
_{a0}dt^{\prime}\label{39}%
\end{equation}
In turn, the solution of this equation is
\begin{align}
\delta\phi & =\frac{\delta\phi^{a}}{2a\left(  t\right)  }\int^{t}\left(
\int^{t^{\prime}}\lambda_{a0}dt^{\prime\prime}\right)  adt^{\prime
}\nonumber\\
& =-\frac{V_{,a}\delta\phi^{a}}{2a\left(  t\right)  }\int^{t}\left(
\int^{t^{\prime}}\left(  \frac{1}{a^{3}}\int^{t^{\prime\prime}}a^{3}%
dt^{\prime\prime\prime}\right)  dt^{\prime\prime}\right)  adt^{\prime
}\label{40}%
\end{align}
One can verify, with some effort, that the $0-0$ and $i=k$ Einstein equations
are automatically satisfied by $\left(  \ref{40}\right)  $ as they should be.
Let us consider a flat universe filled by mimetic dark matter in the presence
of cosmological constant, that is, $V\left(  \phi_{0}^{a}\right)  =V_{0}%
\neq0.$ In this case the solution of Friedmann equations
\begin{equation}
H^{2}=\frac{1}{3}\left(  2\lambda_{0}+V_{0}\right)  ,\text{ \quad}\dot
{H}=-\lambda_{0},\label{41}%
\end{equation}
is
\begin{equation}
a\left(  t\right)  =a_{0}\left(  \sinh\left(  \frac{3}{2}H_{V}t\right)
\right)  ^{2/3}.\label{42}%
\end{equation}
where $H_{V}=\frac{1}{3}V_{0}.$ Substituting this expression in $\left(
\ref{40}\right)  $ we find, after doing the integrals in equation$(\ref{40})$%
\begin{equation}
\delta\phi=-\frac{V_{,a}\delta\phi^{a}}{2V_{0}}t+A\frac{1}{a\left(  t\right)
}\int^{t}a\left(  t^{\prime}\right)  dt^{\prime}+B,\label{43}%
\end{equation}
where $A$ and $B$ are constants of integration. The corresponding
gravitational potential is equal to%
\begin{equation}
\Phi=\delta\dot{\phi}=-\frac{V_{,a}\delta\phi^{a}}{2V_{0}}+A\left(  1-\frac
{H}{a\left(  t\right)  }\int^{t}a\left(  t^{\prime}\right)  dt^{\prime
}\right)  .\label{44}%
\end{equation}
Thus we see that the inhomogenieties in dark energy induce the contribution to
the gravitational potential which do not disappear when the universe turns to
the de Sitter stage. In fact let us separate the contributions from
inhomogeneous dark energy and mimetic dark matter to the gravitational
potential. With this purpose we notice that the contribution of dark energy to
$\Phi$ should vanish at the very early time when the Universe is dominated by
dark matter. Thus we obtain%
\begin{align}
\Phi &  =\frac{\delta V}{3V_{0}}\left(  1-\frac{5}{2}\frac{H_{V}\coth\left(
\frac{3}{2}H_{V}t\right)  }{\sinh^{2/3}\left(  \frac{3}{2}H_{V}t\right)  }%
\int_{0}^{t}\sinh^{2/3}\left(  \frac{3}{2}H_{V}t\right)  dt\right)
\nonumber\\
&  +A_{DM}\left(  1-\frac{H_{V}\coth\left(  \frac{3}{2}H_{V}t\right)  }%
{\sinh^{2/3}\left(  \frac{3}{2}H_{V}t\right)  }\int^{t}\sinh^{2/3}\left(
\frac{3}{2}H_{V}t\right)  dt\right)  ,\label{46}%
\end{align}
where $\delta V=V_{,a}\delta\phi^{a}.$ The first and second terms here
represent the contribution of the dark energy and dark matter to the
gravitational potential respectively. In the limit $H_{V}t\ll1,$ that is at
the time when dark matter dominates over dark energy, this formula simplifies
to
\begin{equation}
\Phi\simeq-\frac{9\delta V}{44V_{0}}\left(  H_{V}t\right)  ^{2}+\frac{3}%
{5}A_{DM},\label{47}%
\end{equation}
where we have skipped the decaying mode. At late time when the universe is
dominated by dark energy, that is, for $H_{V}t\gg1,$ one gets%
\begin{equation}
\Phi\simeq-\frac{\delta V}{2V_{0}}-3A_{DM}e^{-3H_{V}t}.\label{48}%
\end{equation}
In our considerations, the field $\phi^{a}$ which could be responsible for
inhomogeneous dark energy, is fully non-dynamical and it is natural to ask how
it could be generated in the early universe, and what might be the initial
spectrum of the perturbations $\delta\phi^{a}.~$In fact, to generate
$\delta\phi^{a}$ in the early universe we have to modify the second constraint
in $\left(  \ref{4}\right)  .$ The simplest modification of the constraint
$(\ref{4}),$ entirely in the mimetic framework, is to take
\begin{equation}
\frac{\partial\phi^{a}}{\partial x^{\mu}}\frac{\partial\phi}{\partial x^{\nu}%
}g^{\mu\nu}=F^{a}\left(  \phi\right)  ,\label{49}%
\end{equation}
where $F^{a}\left(  \phi\right)  $ are some functions of $\phi$ which we
assume to vanish for large values of $\phi=t$ (in synchronous gauge), that is,
in late Universe. We will skip here the possible justification for this
assumption and concentrate on the question of what is the spectrum in this
case of the generated inhomogenieties of dark energy. For the homogeneous
Friedmann background the constraint $\left(  \ref{49}\right)  $ $\ $becomes%
\begin{equation}
\frac{d\phi_{0}^{a}}{dt}=F^{a}\left(  \phi_{0}\right)  =F^{a}\left(  t\right)
\label{50}%
\end{equation}
and at late times the homogeneous constant field
\begin{equation}
\phi_{0}^{a}=\int F^{a}\left(  t\right)  dt,\label{51}%
\end{equation}
responsible for the dark energy will be generated. In turn, to linear order in
perturbations in the conformal-Newtonian gauge, we obtain
\begin{equation}
\frac{\partial\delta\phi^{a}}{\partial t}=F^{a}\Phi+\frac{\partial F^{a}%
}{\partial\phi}\delta\phi,\label{52}%
\end{equation}
where we have taken into account that $\Phi=\delta\dot{\phi}$ irrespective of
the matter content of the universe. After inflation the gravitational
potential $\Phi$ is time independent at large scales for most of the time and
is equal to $\Phi=\frac{3}{5}A_{DM}$ at the matter dominated stage. Therefore
the generated spectrum of inhomogenieties of $\delta\phi^{a}$ is the same as
the spectrum of $\Phi$ and hence we can set
\begin{equation}
\frac{\delta V}{3V_{0}}=\gamma A_{DM},\label{53}%
\end{equation}
in formula $\left(  \ref{46}\right)  ,$ where $\gamma$ is a numerical
coefficient which can be taken to be of the order of unity and can have either
positive or negative values depending on the choices of $F^{a}$. The formula
$\left(  \ref{46}\right)  $ then becomes
\begin{equation}
\Phi=\left(  1+\gamma\right)  A_{CM}-\left(  1+\frac{5}{2}\gamma\right)
A_{CM}\frac{H}{a\left(  t\right)  }\int^{t}a\left(  t^{\prime}\right)
dt^{\prime}\label{54}%
\end{equation}
The second term above contributes to the late time Sachs-Wolfe effect
(\textit{l}ISW) determined by the integral over the time derivative of the
gravitational potential during transition from matter domination to dark
energy domination epoch \cite{isw}. For $\gamma=-2/5$ the contribution of
\textit{l}ISW vanishes in spite of the presence of dark energy. However,
depending on the sign and value of $\gamma$ \textit{l}ISW can influence the
CMB temperature fluctuations at low multipoles and either decrease or increase
the amplitude of these fluctuations by a significant factor. This opens room
for explanation of low multipole CMB anomalies if they are real. Let us note
that the inhomogenieties in the distribution of dark energy can also influence
the power spectrum at large scales and the formation of structure in the universe.

\subsection*{Acknowledgements}

A.H.C would like to thank the Pauli Center at the Institute for Theoretical
Physics, ETH, Zurich, Switzerland and the Arnold Sommerfeld Institute for
Theoretical Physics at the Ludwig Maximilians University, Munich, Germany for
their hospitality where this research was done. His work is also supported in
part by the National Science Foundation Grant No. Phys-1202671 and
Phys-1518371.. The work of VM is supported by TRR 33 \textquotedblleft The
Dark Universe\textquotedblright\ and the Cluster of Excellence EXC 153
\textquotedblleft Origin and Structure of the Universe\textquotedblright.

\end{document}